\documentclass[]{IEEEphot}
\usepackage{tabularx}
\usepackage{ulem}
\jvol{xx}
\jnum{xx}
\jmonth{XX}
\pubyear{2018}

\begin{document}

\title{Adaptive Spatial Modulation Based Visible Light Communications: SER Analysis and Optimization}

\author{Jin-Yuan Wang$^{1,2}$,~\IEEEmembership{Member,~IEEE}, Jian-Xia Zhu$^{3}$, Sheng-Hong Lin$^{4}$, \\and Jun-Bo Wang$^{2}$,~\IEEEmembership{Member,~IEEE}}
\affil{$^1$College of Telecommunications \& Information Engineering, Nanjing University of Posts and Telecommunications, Nanjing 210003, China.\\
$^2$National Mobile Communications Research Laboratory, Southeast University, Nanjing 211189, China.\\
$^3$Information Technology Research Center, China Telecommunications, Shanghai 201315, China.\\
$^4$School of Automation Engineering, Nanjing Institute of Mechatronic Technology, Nanjing 211135, China.}
\doiinfo{DOI: 10.1109/JPHOT.2018.XXXXXXX\\
1943-0655/\$25.00 \copyright 2018 IEEE}%

\maketitle

\markboth{IEEE Photonics Journal}{Adaptive Spatial Modulation Based VLC: SER Analysis and Optimization}

\begin{receivedinfo}%
Manuscript received XX XX, 2018; revised XX XX, 2018. First published XX XX, 2018.
Current version published XX XX, 2018.
This research was supported by National Natural Science Foundation of China (61571115, 61701254, \& 61602235),
Natural Science Foundation of Jiangsu Province (BK20170901 \& BK20161007),
Key International Cooperation Research Project (61720106003),
the open research fund of National Mobile Communications Research Laboratory, Southeast University (2017D06),
the open research fund of Key Lab of Broadband Wireless Communication and Sensor Network Technology (Nanjing University of Posts and Telecommunications), Ministry of Education (JZNY201706), NUPTSF (NY216009),
the open research fund for Jiangsu Key Laboratory of traffic and transportation security (Huaiyin Institute of Technology) (TTS2017-03),
the open research fund of Key Laboratory of Intelligent Computing \& Signal Processing (Anhui University),
the Open Research Subject of Key Laboratory (Research Base) of Signal and Information Processing, Xihua University (szjj2017-047).
\end{receivedinfo}

\begin{abstract}
Recently, the spatial modulation (SM) technique has been proposed for visible light communication (VLC).
This paper investigates the average symbol error rate (SER) for the VLC using adaptive spatial modulation (ASM).
In the system, the analysis of the average SER is divided into two aspects:
the error probability of the spatial domain and the error probability of the signal domain when the spatial domain is correctly estimated.
Based on the two aspects, the theoretical expression of the average SER is derived.
To further improve the system performance, an optimization problem is proposed to optimize the modulation orders on the LEDs.
The ASM based and the candidate reduction (CR)-ASM based optimization algorithms are proposed to solve the problem, respectively.
Numerical results show that the derived theoretical values of the average SER are quite accurate to evaluate the system performance.
Moreover, compared with the existing schemes, the proposed two algorithms are better choices for VLC.
\end{abstract}


\section{Introduction}
\label{section1}
As a new means of wireless communication, visible light communication (VLC) has drawn much attention recently \cite{BIB00}.
Due to its huge bandwidth, the VLC has become a viable complementary solution to radio frequency wireless communications.
Recently, space modulation has been proposed for future multi-input multi-output (MIMO) VLC \cite{BIB01}.
In space modulation, only a fixed number of light-emitting diodes (LEDs) are active while others are idle at any given time instant.
Therefore, the location-dependent spatial information can be utilized to carry additional information bits to boost the overall spectral efficiency \cite{BIB01_1}.
Compared with the conventional MIMO techniques, the space modulation can avoid the inter channel interference and synchronization problems.

Up to now, some work has been done on space modulation for optical wireless communications (OWC).
As the simplest space modulation technique, the space shift keying (SSK) is first proposed for VLC \cite{BIB02}-\cite{BIB04}.
In SSK, only one LED is active at each time instant and information is conveyed by the spatial constellation diagram.
Although the SSK offers a good power efficiency, its spectral efficiency could be improved by not limiting the number of active LEDs to just one at each time instant.
This leads to the concept of generalized SSK (GSSK) \cite{BIB04_1,BIB05}.
However, in SSK or GSSK, there is no signal modulation.
To utilizing the signal domain information, the spatial modulation (SM) is developed.
In SM, there is also only one active LED at each time instant.
However, in addition to using the spatial constellation diagram, the signal constellation diagram is also employed to convey information.
In \cite{BIB05_1}, the average bit error probability (ABEP) for the SM based OWC under turbulence channels is investigated,
while the ABEP for the SM based VLC is derived in \cite{BIB06}.
In \cite{BIB06_1}-\cite{BIB06_3}, the BERs for the OWC using uncoded SM, coded SM, generalized SM, and multi-stream SM are analyzed, respectively.
Ref. \cite{BIB07} proposes a significant enhancement of optical SM performance by aligning the LEDs and the PDs.
In \cite{BIB08}, the mutual information of the SM based OWC is derived.
In \cite{BIB09}, the SM scheme that combines SSK with pulse position modulation is investigated for OWC.
However, in \cite{BIB06_1}-\cite{BIB09}, the modulation orders on the LEDs are the same as each other.
To improve the system performance, the adaptive modulation schemes should be considered.
In \cite{BIB010}, the channel-adaptive SM is considered for VLC, but the system performance is not analyzed from the theoretical aspect.
Therefore, it is necessary to analyze and optimize the system performance for the VLC using adaptive SM (ASM).

Motivated by \cite{BIB010}, this paper analyzes the average SER performance for ASM based VLC and presents an optimization problem to improve system performance.
The main contributions of this paper are summarized as follows:
\begin{itemize}
  \item Different from \cite{BIB06_1}-\cite{BIB09} using the same modulation orders on all LEDs,
        the ASM scheme for VLC is considered in this paper.
        Moreover, the probability of selecting an LED to transmit information is proportion to the value of the modulation order on the LED.
        In other words, the LED with large modulation order will have a large probability to be selected.
  \item The average SER of the proposed ASM based VLC is analyzed.
        The problem of analyzing the average SER is transformed to the problem of analyzing two error probabilities.
        The first one is the error probability of the spatial domain (i.e., the error probability of estimating the index of the active LED).
        The second one is the error probability of the signal domain when the spatial domain is correctly estimated (i.e., the error probability of estimating the emitting symbol by the active LED when the estimation of the active LED's index is correct).
        Based on the two probabilities, the theoretical expression of the average SER is derived.
  \item To improve the system performance, an optimization problem is proposed to optimize the modulation order on each LED in case of the channel state information (CSI) at the transmitter is well known.
      The ASM based and the candidate reduction (CR)-ASM based optimization algorithms are proposed respectively to solve the problem.
  \item To show the accuracy of the derived theoretical expression and the reasonability of the proposed algorithm,
        all theoretical results are thoroughly confirmed by Monte-Carlo simulations.
\end{itemize}

The remainder of this paper is organized as follows.
In Section \ref{section2}, the system model is described.
In Section \ref{section3}, the theoretical expression of the average SER for the VLC using ASM is analyzed.
Specially, the theoretical expression of SER for VLC using SSK and SMS are shown in Section \ref{section4}.
In Section \ref{section5}, an optimization problem to minimize SER is provided to improve system performance.
Numerical results are presented in Section \ref{section6}.
Finally, Section \ref{section7} draws conclusions.

\section{System Model}
\label{section2}
\begin{figure*}
\centering
\includegraphics[scale=0.6]{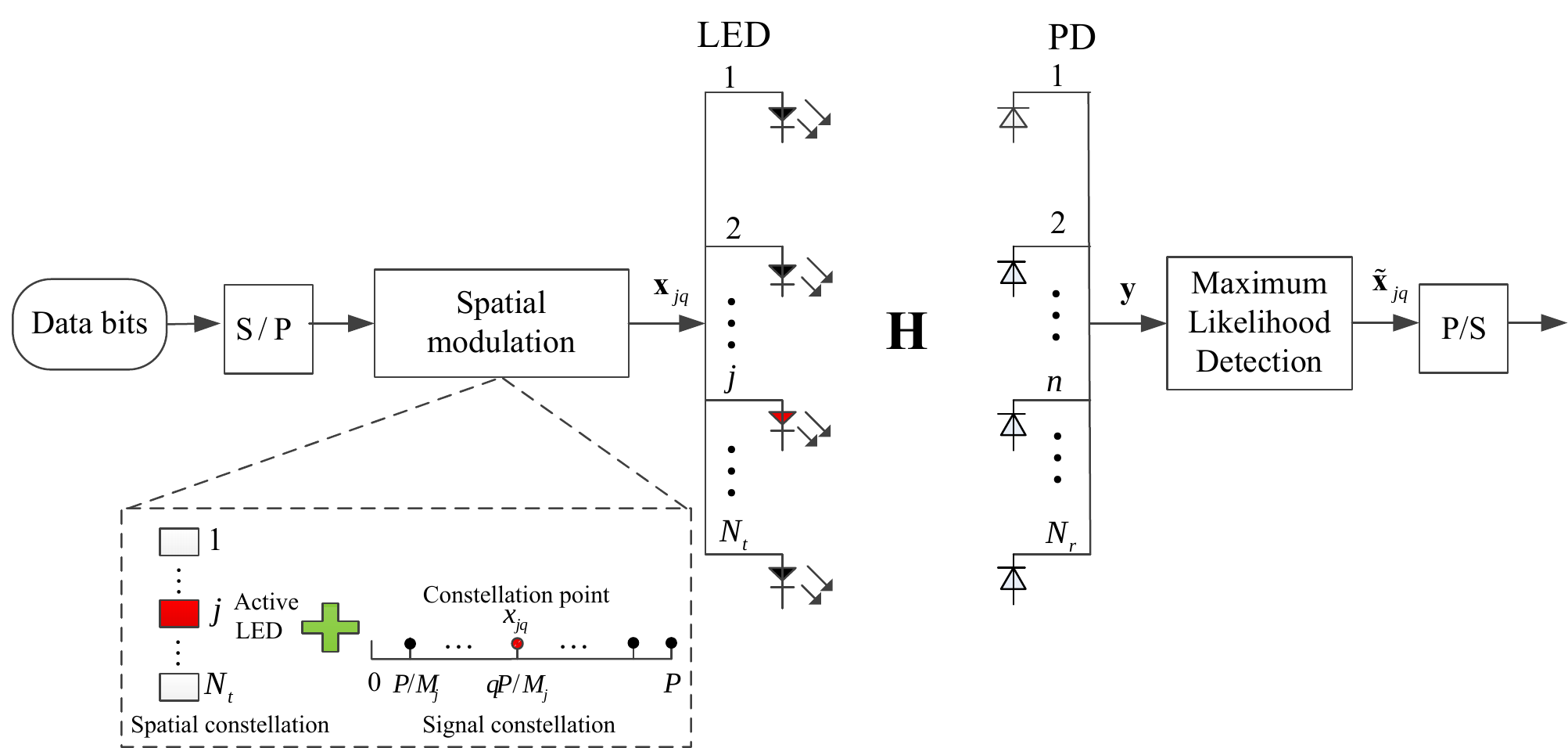}
\caption{The VLC system using ASM }
\label{fig1}
\end{figure*}

Consider a VLC system using ASM, as shown in Fig. \ref{fig1}.
In the system, ${N_{\rm t}}$ LEDs and ${N_{\rm r}}$ PDs are deployed at the transmitter and the receiver, respectively.
Note that the value of ${N_{\rm t}}$ is assumed to be the power of 2.
At the transmitter, only one LED is active at any timeslot and conveys a modulated positive intensity information while the other LEDs are silent.
By using ASM, the information bits are mapped into two parts, where one part indicates the index of the active LED (namely spatial domain bits) and the other part uses the pulse amplitude modulation (PAM) (called signal domain bits).
At the receiver, ${N_{\rm r}}$ PDs receive the optical intensity signals and convert them into electrical signals.

In this paper, the CSI is well known at the transmitter, and the feedback of CSI in uplink can be accomplished by using some feasible wireless technologies, such as wireless radio frequency, visible light, infrared light or millimeter wave \cite{BIBaddA,BIBaddB}. Moreover, the modulation order on each LED can be adaptively updated by using the ASM. Without loss of generality, the modulation order on the $j$-th LED is denoted as ${M_j}$.
If the $j$-th LED is activated and the $q$-th constellation point in ${M_j}$-ary PAM signal constellation is emitted, the $N_{\rm t}$-dimentional transmitted signal vector can be expressed as
\begin{equation}
{\bf{x}}_{jq} = [ 0,\cdots,0, x_{jq},0,\cdots,0]^{\rm{T}}
\label{equ1}
\end{equation}
where ${x_{jq}}$ is the $q$-th signal constellation point in the ${M_j}$-ary constellation diagram.

Therefore, the received signal vector at ${N_{\rm r}}$ PDs can be written as
\begin{eqnarray}
\bf{y} &=& {\bf{H}}{{\bf{x}}_{jq}} +{\bf{ w}}\nonumber \\
 &=& {{\bf{h}}_{j}}{x_{jq}} + {\bf{w}}
\label{equ2}
\end{eqnarray}
where
${\bf{y}} = [ y_1,y_2,\cdots,y_{N_{\rm r}}]^{\rm{T}}$ represents the $N_{\rm r}$-dimensional received signal vector,
and ${\bf{w}} = [ w_1,w_2,\cdots,w_{N_{\rm r}}]^{\rm{T}}$ is the ${N_{\rm r}}$-dimensional noise vector.
Each element in ${\bf{w}}$ can be modeled as independent real valued additive white Gaussian noise with zero mean and variance ${\sigma ^2}$. ${\bf{H}} = [{\bf{h}}_1,{\bf{h}}_2, \cdots ,{\bf{h}}_j, \cdots ,{\bf{h}}_{{N_{\rm t}}}]$ is the $( {N_{\rm r}} \times {N_{\rm t}} )$-dimensional channel matrix, which can be expressed as
\begin{equation}
{\bf{H}} = \left[ {\begin{array}{*{20}{c}}
{{h_{11}}}&{{h_{12}}}& \cdots &{{h_{1{N_{\rm t}}}}}\\
{{h_{21}}}&{{h_{22}}}& \cdots &{{h_{2{N_{\rm t}}}}}\\
 \vdots & \vdots & \ddots & \vdots \\
{{h_{{N_{\rm r}}1}}}&{{h_{{N_{\rm r}}2}}}& \cdots &{{h_{{N_{\rm r}}{N_{\rm t}}}}}
\end{array}} \right]
\label{equ3}
\end{equation}
where ${h_{i,j}}$ represents the channel gain of the VLC link between the $j$-th LED and the $i$-th PD in indoor line of sight environment, and it can be written as \cite{BIB0101}
\begin{equation}
{h_{ij}} = \left\{ {\begin{array}{*{20}{c}}
{\frac{{\left( {k + 1} \right)A}}{{2\pi d_{ij}^2}}{{\cos }^k}\left( {{\phi _{ij}}} \right)\cos \left( {{\varphi _{ij}}} \right),}&{0 \le {\varphi _{ij}} \le {\Psi _c}}\\
{0,}&{{\varphi _{ij}} > {\Psi _c}}
\end{array}} \right.
\label{equ4}
\end{equation}
where $k$ is the Lambertian emission order given as $k =  - \ln 2/\ln \left( {\cos {\Phi _{1/2}}} \right)$,
$A$ is the physical area of the PD,
${\Phi _{1/2}}$ is the semi-angle at half-power of the LED,
${d_{ij}}$ is the transmission distance between the $j$-th LED and the $i$-th PD,
${\phi _{ij}}$ and ${\varphi _{ij}}$ are the angle of emission and incidence from the $j$-th LED to the $i$-th PD,
${\Psi _c}$ denotes the field of view of the PD.

At the receiver, the maximum-likelihood principle is used to detect the received signal, which can be expressed as
\begin{eqnarray}
[\tilde j,\tilde q]\!\!\!\!&=&\!\!\!\! \arg \mathop {\max }\limits_{1 \le j \le {N_{\rm t}},1 \le q \le {M_j}} {\kern 1pt} {\kern 1pt} {f_{{{\bf{y}}|{{\bf{x}}_{jq}},{\bf{H}}}}}\left( {{\bf{y}}|{{\bf{x}}_{jq}},{\bf{H}}} \right)\ \nonumber \\
 &=&\!\!\!\! \arg \mathop {\min {\kern 1pt} }\limits_{1 \le j \le {N_{\rm t}},1 \le q \le {M_j}} \left\| {{\bf{y}} - {\bf{H}}{{\bf{x}}_{jq}}} \right\|_{\rm F}^2\nonumber \\
 &=& \!\!\!\!\arg \mathop {\min {\kern 1pt} {\kern 1pt} }\limits_{1 \le j \le {N_{\rm t}},1 \le q \le {M_j}} \!\! \left\{\! {\left\| {{{\bf{h}}_j}{x_{jq}}} \right\|_{\rm F}^2 \!-\! 2({{\bf{y}}^T}{{\bf{h}}_j}{x_{jq}})} \!\right\}
\label{equ5}
\end{eqnarray}
where $\tilde j$ is the estimated index of the active LED, $\tilde q$ is the estimated signal constellation point of the ${M_{\tilde j}}$-ary PAM constellation, ${f_{{{\bf{y}}|{{\bf{x}}_{jq}},{\bf{H}}}}}\left( {{\bf{y}}|{{\bf{x}}_{jq}},{\bf{H}}} \right)$ is the conditional probability density function of ${\bf{y}}$ in the condition of signal vector ${{\bf{x}}_{jq}}$ and channel matrix ${\bf{H}}$. ${\left\|  \cdot  \right\|_{\rm F}}$ represents the Frobenius norm of a vector or matrix.

\section{Average SER Analysis for ASM Based VLC}
\label{section3}
Based on the system model, the error performance of the ASM based VLC system will be analyzed in this section.

Define $\Pr \left( j \right)$ be the probability of selecting the $j$-th LED to transmit information.
In this paper, the ASM scheme is employed.
Therefore, the LED with good CSI will be assigned large modulation order.
Such an LED will be selected with high probability to transmit information.
That is, if the modulation order on an LED is larger, the probability of selecting the LED to transmit information will be larger.
According to the proportional selection scheme, $\Pr \left( j \right)$ is modeled by \cite{BIB0102}
\begin{equation}
\Pr (j){\kern 1pt} {\kern 1pt}  = \frac{{{M_j}}}{{\sum\limits_{i = 1}^{{N_{\rm t}}} {{M_i}} }}, \; j = 1,2, \cdots ,{N_{\rm t}}
\label{equ6}
\end{equation}

Define ${P_e}$ be the average SER of the VLC system, and ${P_{ej}}$ be the SER of selecting the $j$-th LED to transmit information.
In other words, when the $j$-th LED is selected as the active LED at the transmitter, the symbol error rate at the receiver is ${P_{ej}}$.
Therefore, the average SER \uline{${P_e}$} is given by
\begin{eqnarray}
{P_e} &=& \sum\limits_{j = 1}^{{N_{\rm t}}} {\Pr (j){P_{ej}}} \nonumber \\
  &=& \sum\limits_{j = 1}^{{N_{\rm t}}} {\frac{{{M_j}}}{{\sum\limits_{i = 1}^{{N_{\rm t}}} {{M_i}} }}{P_{ej}}}
\label{equ7}
\end{eqnarray}
When the $j$-th LED is activated to convey information at the transmitter, ${P_{ej}}$ can be divided into three parts:
(a) the spatial domain information is wrong and the signal domain information is correct; (b) the spatial domain information is wrong and the signal domain information is wrong; (c) the spatial domain information is correct and the signal domain information is wrong.
Assume that the error probability of spatial domain information is $P_{aj}$ and the conditional error probability of signal domain information when the spatial domain is correctly estimated is $P_{sj}$.
Therefore, the error probabilities for parts (a), (b) and (c) are given by ${P_{aj}}(1 - {P_{sj}})$, ${P_{aj}}{P_{sj}}$ and $(1 - {P_{aj}}){P_{sj}}$, respectively. Therefore, ${P_{ej}}$ can be expressed as
\begin{eqnarray}
{P_{ej}} &=& {P_{aj}}(1 - {P_{sj}}) + {P_{aj}}{P_{sj}} + (1 - {P_{aj}}){P_{sj}} \nonumber\\
 &=& {P_{aj}} + (1 - {P_{aj}}){P_{sj}}
\label{equ8}
\end{eqnarray}
Therefore, the problem of analyzing the average SER is transformed into the problem of analyzing ${P_{aj}}$ and ${P_{sj}}$.
In following two subsections, the two error probabilities will be analyzed respectively.

\subsection{Analysis of ${P_{sj}}$ in signal domain}
\label{section3_1}
In VLC, the information is modulated as the instantaneous optical intensity, and thus the transmit optical signal should be non-negative.
In addition, the instantaneous optical intensity of an LED is constrained by its peak optical intensity.
Therefore, we have
\begin{equation}
{\rm{0}} \le {x_{jq}} \le P,\;\forall j,q
\label{equ9}
\end{equation}
where $P$ is the allowed peak optical intensity of each LED.

\begin{figure}
\centering
\includegraphics[scale=0.6]{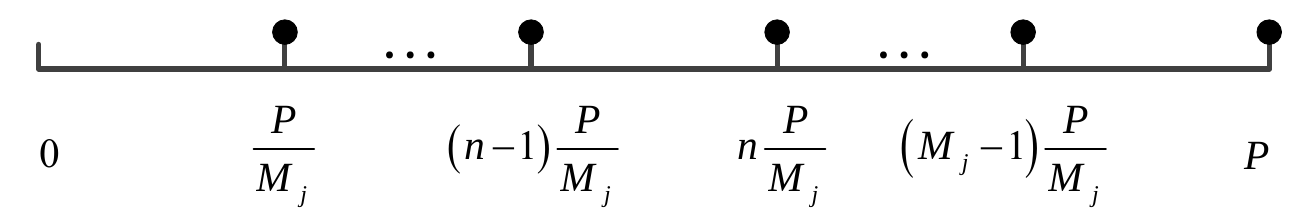}
\caption{ The signal constellation diagram for the ${M_j}$-ary PAM\label{fig2}}
\end{figure}

In this paper, the ${M_j}$-ary PAM is employed by the $j$-th LED, and the signal constellation diagram is given in Fig. \ref{fig2}.
In the figure, the signal space is $\left[ {0,P} \right]$. The constellation points are supposed to be equally spaced, and the optical intensity $nP/M_j,\;n = 1,2, \cdots ,{M_j}$ are the constellation points while intensity zero is not.
Therefore, the spacing between any two adjacent constellation points is given by
\begin{equation}
{A_j} = \frac{P}{{{M_j}}},\;j = 1,2, \cdots ,{N_{\rm t}}
\label{equ10}
\end{equation}

\begin{figure}
\centering
\includegraphics[scale=0.5]{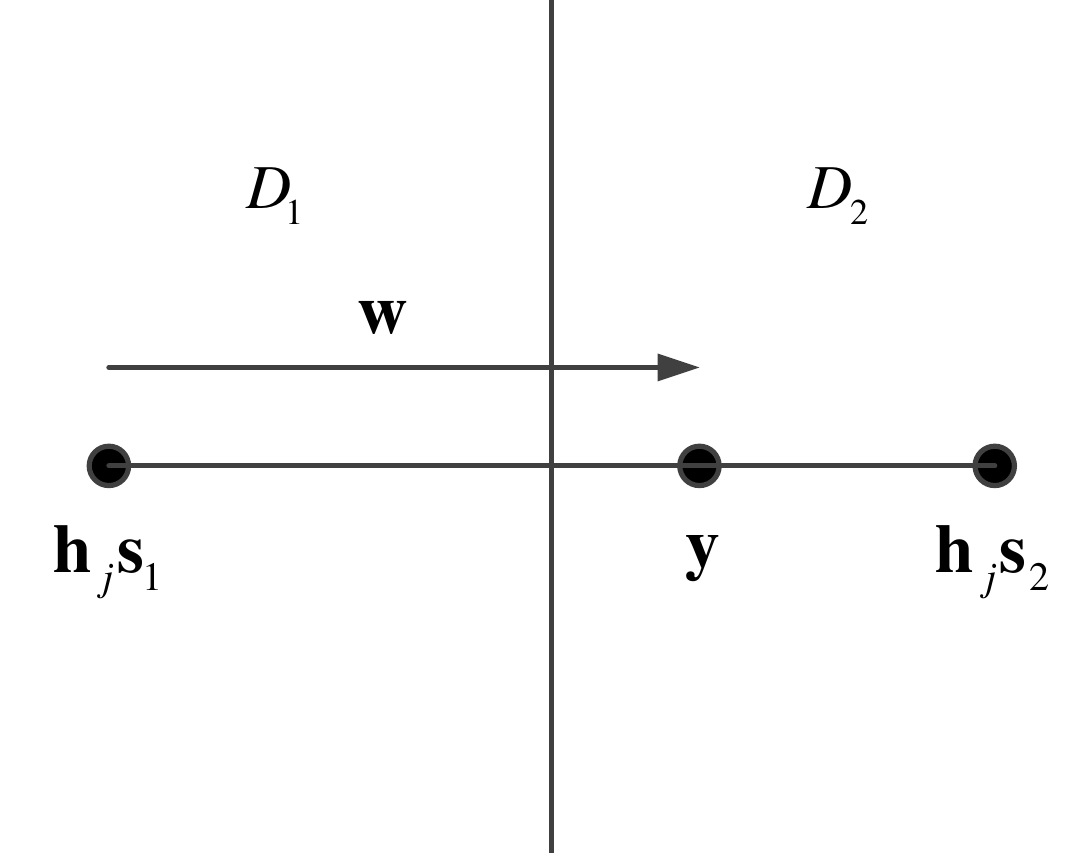}
\caption{ Decision regions for two adjacent signal vectors when the $j$-th LED is selected\label{fig3}}
\end{figure}

Let ${{\bf{s}}_1}$ and ${{\bf{s}}_2}$ be any two adjacent emitting signal vectors in constellation space,
and they have equal probabilities. Therefore, the two decision regions can be determined by using the nearest-neighbor principle.
When the $j$-th LED is selected, the decision regions (i.e., ${D_1}$ and ${D_2}$) are shown in Fig. \ref{fig3}.
If the constellation point ${{\bf{s}}_1}$ is sent, an error occurs if ${\bf{y}}$ is located in ${D_2}$.
Therefore, the error probability is given by \cite{BIB011}
\begin{eqnarray}
{P_b} &=& \Pr \left[ {{\rm{error}}|j{\rm{ - th}}\;{\rm{LED}}\;{\rm{is}}\;{\rm{selected}},\;{{\bf{s}}_1}{\kern 1pt} {\kern 1pt} {\rm{is}}\;{\rm{sent}}} \right] \nonumber \\
&=& \Pr \left[ {{\bf{w}} \cdot \left( {{{\bf{h}}_j}{{\bf{s}}_2} - {{\bf{h}}_j}{{\bf{s}}_1}} \right) > \frac{{d_{12}^2}}{2}} \right]
\label{equ11}
\end{eqnarray}
where ${d_{12}} = {\left\| {{{\bf{h}}_j}{{\bf{s}}_2} - {{\bf{h}}_j}{{\bf{s}}_1}} \right\|_{\rm F}}$, ${{\bf{w}} \cdot \left( {{{\bf{h}}_j}{{\bf{s}}_2} - {{\bf{h}}_j}{{\bf{s}}_1}} \right)}$ is a real Gaussian random variable with zero-mean and variance ${\sigma ^2}\left\| {{{\bf{h}}_j}{{\bf{s}}_2} - {{\bf{h}}_j}{{\bf{s}}_1}} \right\|_{\rm F}^2$.

Therefore, ${P_b}$ can be further written as
\begin{eqnarray}
{P_b} 
&=& {\cal Q}\left( {\frac{{{d_{12}}}}{{2\sigma }}} \right)
\label{equ12}
\end{eqnarray}
where ${\cal Q}(x) = \int_x^\infty  {{{{e^{ - {t^2}/2}}} \mathord{\left/
 {\vphantom {{{e^{ - {t^2}/2}}} {\sqrt {2\pi } }}} \right.
 \kern-\nulldelimiterspace} {\sqrt {2\pi } }}} {\rm{d}}t$ denotes the Gaussian $\cal Q$-function.

When the $j$-th LED is activated, the minimum distance between the received signals can be expressed as
\begin{eqnarray}
d_{\min }^j &=& {\left\| {{{\bf{h}}_j}{A_j}} \right\|_{\rm F}} = {\left\| {{{\bf{h}}_j}\frac{P}{{{M_j}}}} \right\|_{\rm F}} \nonumber \\
 &=& \frac{P}{{{M_j}}}{\left\| {{{\bf{h}}_j}} \right\|_{\rm F}} = \frac{P}{{{M_j}}}\sqrt {\sum\limits_{n = 1}^{{N_{\rm r}}} {h_{nj}^2} }
\label{equ13}
\end{eqnarray}

In the ${M_j}$-ary PAM constellation, there are ${M_j} - 2$ inner points and 2 outer points.
For the outer points, the error probability is one-half of the error probability of an inner point because of noise causing error in only one direction.
Therefore, ${P_{sj}}$ is given by \cite{BIB011}
\begin{eqnarray}
{P_{sj}}\!\!\!\!&=&\!\!\!\! \frac{1}{{{M_j}}}\!\!\sum\limits_{m = 1}^{{M_j}}\!\! {\Pr \left[ {{\rm error}|m- {\rm th \;constellation \; point \; is \;sent}} \right]} \nonumber \\
 &=&\!\!\!\! \frac{1}{{{M_j}}}\left[ {2\left( {{M_j} - 2} \right){\cal Q}\left( {\frac{{d_{\min }^j}}{{2\sigma }}} \right) + 2{\cal Q}\left( {\frac{{d_{\min }^j}}{{2\sigma }}} \right)} \right] \nonumber \\
 &=&\!\!\!\! \frac{{2\left( {{M_j} - 1} \right)}}{{{M_j}}}{\cal Q}\left( {\frac{{P\sqrt {\sum\limits_{n = 1}^{{N_{\rm r}}} {h_{nj}^2} } }}{{2\sigma {M_j}}}} \right)
\label{equ14}
\end{eqnarray}

\subsection{Analysis of ${P_{aj}}$ in spatial domain}
\label{section3_2}
When the $j$-th LED is selected, let ${S_j},j = 1,2, \cdots ,{N_{\rm t}}$ stand for the set of the transmit constellation points, which can be written as
\begin{equation}
{S_j} = \left\{ {\frac{P}{{{M_j}}},\frac{{2P}}{{{M_j}}},\frac{{3P}}{{{M_j}}}, \cdots ,P} \right\},j = 1,2, \cdots ,{N_{\rm t}}
\label{equ15}
\end{equation}

Moreover, If the $j$-th LED is activated and the ${x_{jq}}$ is emitted, the minimum distance between ${x_{jq}}$ and the other signal transmitted by different LED can be expressed as
\begin{equation}
{D_{jq}} = \mathop {\min }\limits_{{\kern 1pt} {\kern 1pt} {\kern 1pt} {\kern 1pt} {\kern 1pt} {\kern 1pt} {x_{ip}} \in {S_i}1 \le i \le {N_{\rm t}}.{\kern 1pt} i \ne j{\kern 1pt} {\kern 1pt} {\kern 1pt} {\kern 1pt} {\kern 1pt} {\kern 1pt} {\kern 1pt} {\kern 1pt} {\kern 1pt} {\kern 1pt} {\kern 1pt} } {\left\| {{{\bf{h}}_j}{x_{jq}} - {{\bf{h}}_i}{x_{ip}}} \right\|_{\rm F}}
\label{equ16}
\end{equation}

At high SNR, if the estimation of the spatial domain of the ${x_{jq}}$ is error, it is most likely to be detected as the signal which is transmitted by different LED and is the nearest to the ${x_{jq}}$. So the error probability of the spatial domain of the ${x_{jq}}$ can be expressed as
\begin{equation}
{P_{jq}} = {\cal Q}\left( {\frac{{{D_{jq}}}}{{2\sigma }}} \right)
\label{equ17}
\end{equation}

Furthermore, the error probability of the active LED index $j$ can be expressed as the average error probability of the all signals transmitted by the $j$-th LED. Finally, ${P_{aj}}$ can be written as
\begin{eqnarray}
{P_{aj}} = \sum\limits_{q = 1}^{{M_j}} {\frac{1}{{{M_j}}}} {P_{jq}} = \sum\limits_{q = 1}^{{M_j}} {\frac{1}{{{M_j}}}} {\cal Q}\left( {\frac{{{D_{jq}}}}{{2\sigma }}} \right)
\label{equ18}
\end{eqnarray}

\subsection{Theoretical expression of average SER ${P_e}$}
\label{section3_3}
Using (\ref{equ14}) and (\ref{equ18}) into (\ref{equ8}), we have
\begin{eqnarray}
{P_{ej}}
 = \sum\limits_{q = 1}^{{M_j}} {\frac{1}{{{M_j}}}} {\cal Q}\left( {\frac{{{D_{jq}}}}{{2\sigma }}} \right) + \left[ {1 - \sum\limits_{q = 1}^{{M_j}} {\frac{1}{{{M_j}}}} {\cal Q}\left( {\frac{{{D_{jq}}}}{{2\sigma }}} \right)} \right] \frac{{2\left( {{M_j} - 1} \right)}}{{{M_j}}}{\cal Q}\left( {\frac{{P{{ \sqrt{\sum\limits_{n = 1}^{{N_{\rm r}}} {{h^2_{nj}}} }}}}}{{2\sigma {M_j}}}} \right)
\label{equ19}
\end{eqnarray}

Finally using (\ref{equ19}) into (\ref{equ7}), the theoretical expression of the SER for the ASM based VLC can be derived as
\begin{equation}
{P_e} = \sum\limits_{j = 1}^{{N_{\rm t}}} {\frac{{{M_j}}}{{\sum\limits_{i = 1}^{{N_{\rm t}}} {{M_i}} }}\!\!\left\{ {\sum\limits_{q = 1}^{{M_j}} {\frac{1}{{{M_j}}}} {\cal Q}\left( {\frac{{{D_{jq}}}}{{2\sigma }}} \right) \!+\! \frac{{2\left( {{M_j} \!-\! 1} \right)}}{{{M_j}}}\!\!\left[\! {1 \!-\! \sum\limits_{q = 1}^{{M_j}} {\frac{1}{{{M_j}}}} {\cal Q}\left( {\frac{{{D_{jq}}}}{{2\sigma }}} \right)}\! \right]\!\! {\cal Q}\!\left( {\frac{{P{\sqrt{ {\sum\limits_{n = 1}^{{N_{\rm r}}} {{h^2_{nj}}} } }}}}{{2\sigma {M_j}}}} \right)} \right\}}
\label{equ20}
\end{equation}

\section{Average SER Analysis for Some Special Cases}
\label{section4}
For the ASM, both the signal constellation and the spatial constellation are considered. As is known, the SSK and the SM with the same modulation orders (named as SMS) are the special cases of the ASM. In this section, the average SER for these special cases will be investigated, respectively.
\subsection{SSK}
\label{section4_1}
In SSK, only the spatial constellation is employed to convey information bits.
In this case, the error probability for the SSK only includes the error of estimating the index of the active LED.
Therefore, according to (\ref{equ20}), the theoretical expression of the average SER for SSK based VLC can be derived as
\begin{eqnarray}
{P_{e,SSK}} = \sum\limits_{j = 1}^{{N_{\rm t}}} {\frac{1}{{{N_{\rm t}}}}\sum\limits_{q = 1}^{{M_j}} {\frac{1}{{{M_j}}}{\cal Q}\left( {\frac{{{D_{jq}}}}{{2\sigma }}} \right)} }
\label{equ21}
\end{eqnarray}

\subsection{SMS}
\label{section4_2}
In SMS, both the signal constellation and the spatial constellation are considered.
However, the modulation orders on LEDs in the signal constellations are the same as each other.
Without loss of generality, let the modulation order on each LED be $M$.
According to (\ref{equ20}), the theoretical expression of the average SER for the SMS based VLC can be derived as
\begin{equation}
{P_{e,SMS}} \!=\! \sum\limits_{j = 1}^{{N_{\rm t}}} {\frac{1}{{{N_{\rm t}}}}\!\!\left\{ {\sum\limits_{q = 1}^{{M}} {\frac{1}{{{M}}}} {\rm Q}\left( {\frac{{{D_{jq}}}}{{2\sigma }}} \right) \!+\! \frac{{2\left( {M \!-\! 1} \right)}}{M}\left[ {1 \!-\! \sum\limits_{q = 1}^{{M}} {\frac{1}{{{M}}}} {\cal Q}\left( {\frac{{{D_{jq}}}}{{2\sigma }}} \right)} \right]\! {\cal Q}\!\!\left( {\frac{{P{\sqrt{ {\sum\limits_{n = 1}^{{N_{\rm r}}} {{{ {{h^2_{nj}}}}}} }}}}}{{2\sigma M}}} \right)}\! \right\}}
\label{equ22}
\end{equation}

\section{Modulation Order Optimization}
\label{section5}
\subsection{Problem formulation and solving}
In (\ref{equ20}), the average SER performance can be improved by selecting an appropriate modulation order combination on the LEDs.
For the SM based VLC, the spectrum efficiency can be expressed as
\begin{eqnarray}
m = {\log _2}\left( {{N_{\rm t}}} \right) + \frac{1}{{{N_{\rm t}}}}\log_2 \left( {\mathop \Pi \limits_{j = 1}^{{N_{\rm t}}} {M_j}} \right)
\label{equ23}
\end{eqnarray}

When the CSI is well known at the transmitter, this section aims to optimize the modulation orders on the LEDs by minimizing the average SER under the condition of fixed spectrum efficiency. Mathematically, the optimization problem for the ASM based VLC can be written as
\begin{eqnarray}
&&[{{\tilde M}_1},{{\tilde M}_2}, \ldots ,{{\tilde M}_{{N_{\rm t}}}}] = \mathop {\min }\limits_{{M_j},j = 1, \cdots ,{N_{\rm t}}} P_e \nonumber \\
{\rm{s}}{\rm{.t}}{\rm{.}}&& m = {\log _2}\left( {{N_{\rm t}}} \right) + \frac{1}{{{N_{\rm t}}}}\log_2 \left( {\mathop \Pi \limits_{j = 1}^{{N_{\rm t}}} {M_j}} \right)
\label{equ24}
\end{eqnarray}
In optimization problem (\ref{equ24}),
the minimum value of the objective is the global minimum,
and the corresponding modulation order combination is the optimal solution.
However, the problem (\ref{equ24}) is an integer optimization problem, which is very complex and non-convex.
Consequently, it is very hard to derive the optimal solution.
In this paper, an ASM scheme is proposed to solve problem (\ref{equ24}), which is shown in Fig. \ref{fig4}.
\begin{figure}[h!]
\hrulefill\\
\textbf{Algorithm 1:} (The ASM scheme)\\
\textbf{Step 1):} Given the positions of the LEDs and the PDs to obtain the channel gain matrix ${\bf{H}}$, and given the spectrum efficiency of VLC system;\\
\textbf{Step 2):} Find all modulation order combinations ${\bf D} = \left\{ {{d_1},{d_2},...,{d_L}} \right\}$, where $L$ is the total number of possible combinations and ${d_i} = \left[ {M_1^i,M_2^i, \cdots ,M_{{N_{\rm t}}}^i} \right]$ denotes the $i$-th modulation order combination, $M_j^i$ denotes the modulation order of the $j$-th LED in the $i$-th modulation order combination;\\
\textbf{Step 3):} Compute the theoretical value of the average SER by using (\ref{equ20}) for all combinations in $\bf{D}$;\\
\textbf{Step 4):} Select the combination with the minimum average SER as the output.\\
\vspace*{2pt} \hrulefill
\caption{The ASM scheme}
\label{fig4}
\end{figure}

Note that the proposed ASM scheme is an exhaust searching algorithm, which has very high computational complexity.
Referring to \cite{BIB019}, a candidate reduction ASM (CR-ASM) scheme is employed in this paper to shrink the search space, which can provide considerable complexity reduction.
Similarly, define ${\bf D} = \left\{ {{d_1},{d_2},...,{d_L}} \right\}$ is the set of all possible modulation order combinations,
where ${d_i} = \left[ {d_1^i,d_2^i, \cdots ,d_{N_{\rm t}}^i} \right] = \left[ {M_1^i,M_2^i, \cdots ,M_{{N_{\rm t}}}^i} \right]$ denotes the $i$-th modulation order combination.
For the $i$-th modulation order combination $d_i$, the variance of this combination is defined as
\begin{equation}
V\left( {{d_i}} \right) = \frac{1}{{{N_{\rm t}}}}\sum\limits_{n = 1}^{{N_{\rm t}}} {{{\left( {d_n^i - {{\bar d}^i}} \right)}^2}}
\label{equ25}
\end{equation}
where ${\bar d^i}$ is the mean of all elements in ${d_i}$.

It is shown in \cite{BIB019} that the value of $V$ is a good metric to classify the legitimate candidates.
The candidates are classified according to their variances and probabilities of occurrence, to derive the CR-ASM scheme. By using the CR-ASM scheme, the searching space $\bf{D}$ reduces to ${\bf D}_r$, which is limited to the candidates with the smallest and the second smallest $V$ values in (\ref{equ25}).
That is because the other candidate cases have little effect on system performance.
To facilitate the understanding, the stepwise procedures of the CR-ASM scheme are shown in Fig. \ref{fig4_1}.
\begin{figure}[!h]
\hrulefill\\
\textbf{Algorithm 2:} (The CR-ASM scheme)\\
\textbf{Step 1):} Given the positions of the LEDs and the PDs to obtain channel gain matrix ${\bf{H}}$, and given the spectrum efficiency of VLC system;\\
\textbf{Step 2):} Find all candidates ${\bf D} = \left\{ {{d_1},{d_2},...,{d_L}} \right\}$, where $L$ is the total number of possible combinations and ${d_i} = \left[ {d_1^i,d_2^i, \cdots ,d_{{N_t}}^i} \right]$ denotes the $i$-th candidates, $d_j^i$ denotes the information bits conveyed by the $j$-th LED in the $i$-th candidate; \\
\textbf{Step 3):} Compute the $V$ values by using (\ref{equ25}) for all combinations in $\bf{D}$;\\
\textbf{Step 4):} Obtain the searching space ${\bf D}_r$ by selecting the combinations with the smallest and the second smallest $V$ values in (\ref{equ25}); \\
\textbf{Step 5):} Compute the theoretical value of the average SER by using (\ref{equ20}) for all combinations in ${\bf D}_r$;\\
\textbf{Step 6):} Select the combination with the minimum average SER as the output.\\
\vspace*{2pt} \hrulefill
\caption{The CR-ASM scheme}
\label{fig4_1}
\end{figure}

\subsection{Time complexity analysis}
In this subsection, the time complexities of the two schemes will be analyzed.
Here, the time complexity of an algorithm is the running time expressed as a function of the size of the input parameters,
which can be expressed by using the big-$O$ notation.
For simplicity, only main computational blocks of the proposed schemes are considered to evaluate the time complexity.
That is, the time complexity is measured based on the numbers of multiplication and addition in main computational blocks.

For the ASM scheme, the numbers of multiplication and addition $O(L{N_{\rm r}}N_{\rm t}^2 + L{N_{\rm t}}\sum\nolimits_{j = 1}^{{N_{\rm t}}} {{M_j}} )$ and $O(L{N_{\rm r}}{N_{\rm t}} + L\sum\nolimits_{j = 1}^{{N_{\rm t}}} {{M_j}} )$, respectively.
For CR-ASM scheme, the numbers of multiplication and addition are $O({N_{\rm r}}N_{\rm t}^2 + L{N_{\rm t}}{\rm{ + }}{N_{\rm t}}\sum\nolimits_{j = 1}^{{N_{\rm t}}} {{M_j}} )$  and $O({N_{\rm r}}{N_{\rm t}} + L + \sum\nolimits_{j = 1}^{{N_{\rm t}}} {{M_j}} )$, respectively.
Therefore, both the ASM and CR-ASM schemes are time-efficient algorithms.
Moreover, the complexity of the CR-ASM scheme is lower than that of the ASM scheme, which indicates that the CR-ASM scheme is a better choice for practical VLC systems.

\section{Numerical Results}
\label{section6}
In this section, some classical results will be shown to verify the accuracy of the theoretical value of the average SER.
Moreover, the performance of the proposed ASM and CR-ASM schemes will be shown.
For comparison, the performance of the SSK scheme and the SMS scheme in section \ref{section4} will also be shown.
In the simulation, the room size is set to be $5{\rm{m}} \times 4{\rm{m}} \times 3{\rm{m}}$. Four different scenarios of $2 \times 1$ (i.e., ${N_{\rm t}} = 2$ and ${N_{\rm r}} = 1$), $4 \times 1$, $2 \times 2$ and $4 \times 4$ VLC systems are considered, and the locations of the LEDs and the PDs are shown in Table \ref{tab1}. Other simulation parameters are presented in Table \ref{tab2}.

\begin{table*}[h!]
\caption{Locations of LEDs and PDs for four simulation scenarios}
\begin{center}
\begin{tabularx}{11cm}{|p{2cm}|X|X|}\hline\hline
\centering \textbf{Scenarios}    &\centering \textbf{Location of LEDs}  &\centering \textbf{Location of PDs}
\tabularnewline\hline
\centering Scenario 1 &\centering (1.8, 2.0, 3.0), (1.8, 3.0, 3.0)  &\centering (2.0, 1.5, 0.8)
\tabularnewline\hline
\centering Scenario 2 &\centering (1.8, 1.0, 3.0), (1.8, 3.0, 3.0), (3.2, 1.0, 3.0), (3.2, 3.0, 3.0) &\centering (2.0, 1.5, 0.8)
\tabularnewline\hline
\centering Scenario 3 &\centering (1.5, 1.0, 3.0), (1.5, 3.5, 3.0)  &\centering (1.5, 1.0, 0.8), (2.0, 2.5, 0.8)
\tabularnewline\hline
\centering Scenario 4 &\centering (1.5, 1.0, 3.0), (1.5, 3.0, 3.0), (3.5, 1.0, 3.0), (3.5, 3.0, 3.0) &\centering (2.0, 2.0, 0.8), (2.0, 3.0, 0.8), (1.5, 3.0, 0.8), (1.5, 2.0, 0.8)
\tabularnewline\hline\hline
\end{tabularx}
\end{center}
\label{tab1}
\end{table*}

\begin{table}[!h]
\caption{Main simulation parameters}
\begin{center}
\begin{tabularx}{8cm}{|p{3.5cm}|p{1.5cm}|X|}\hline\hline
\centering \textbf{Parameters}    &\centering \textbf{Symbols}  &\centering \textbf{Values}
\tabularnewline\hline
\centering Semiangle at half power &\centering ${\Phi _{1/2}} $    &\centering 35 deg
\tabularnewline\hline
\centering Detector area &\centering $A$ &\centering 1.0 ${\rm{c}}{{\rm{m}}^2}$
\tabularnewline\hline
\centering Receiver FOV(half angle) &\centering ${\Psi _c}$ &\centering 60 deg
\tabularnewline\hline\hline
\end{tabularx}
\end{center}
\label{tab2}
\end{table}

\begin{figure}
\begin{minipage}[t]{0.5\linewidth}
\centering
\includegraphics[width=2.7in]{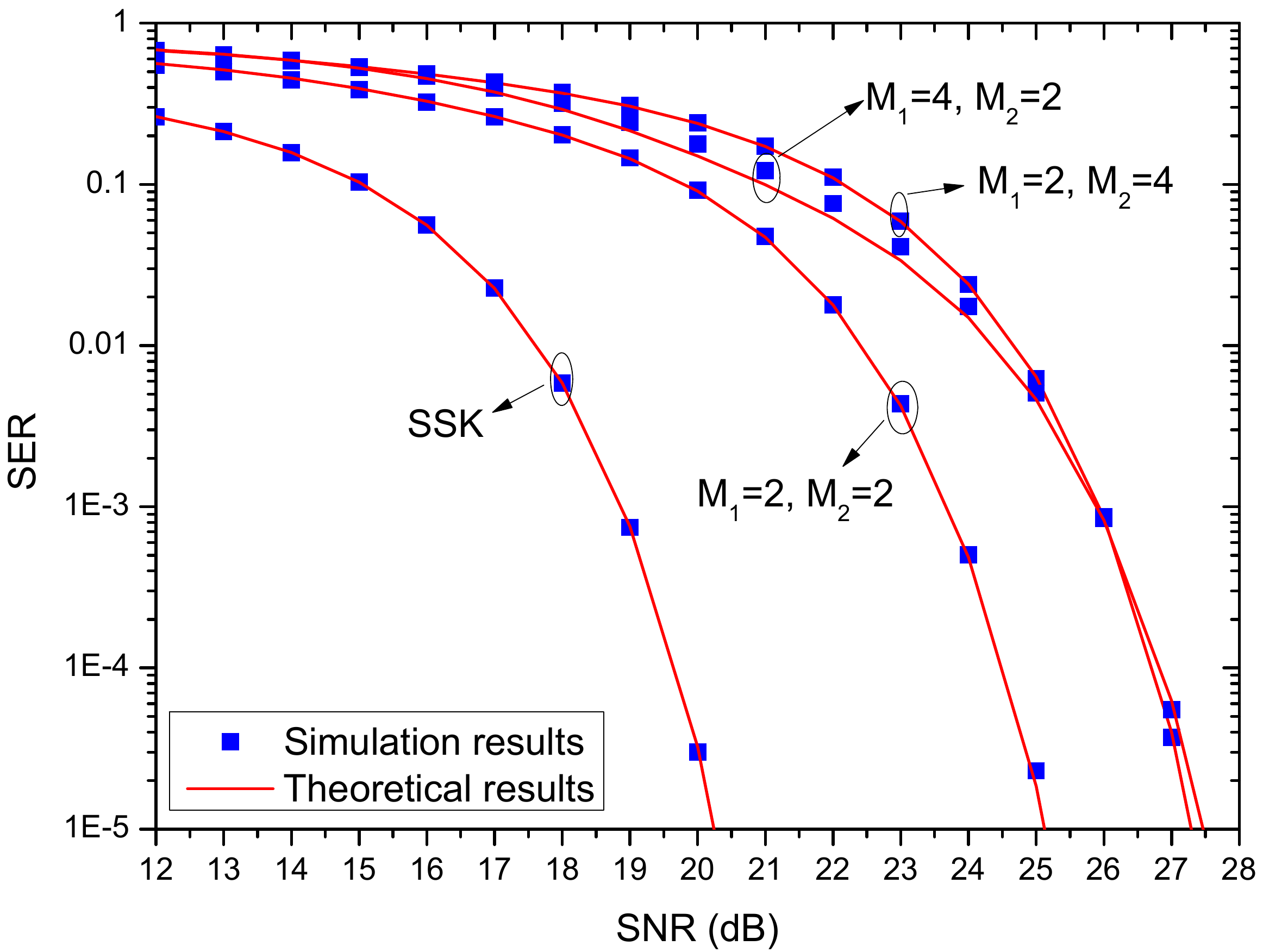}
\caption{SER versus SNR for the $2 \times 1$ VLC system under scenario 1}
\label{fig5}
\end{minipage}
\begin{minipage}[t]{0.5\linewidth}
\centering
\includegraphics[width=2.7in]{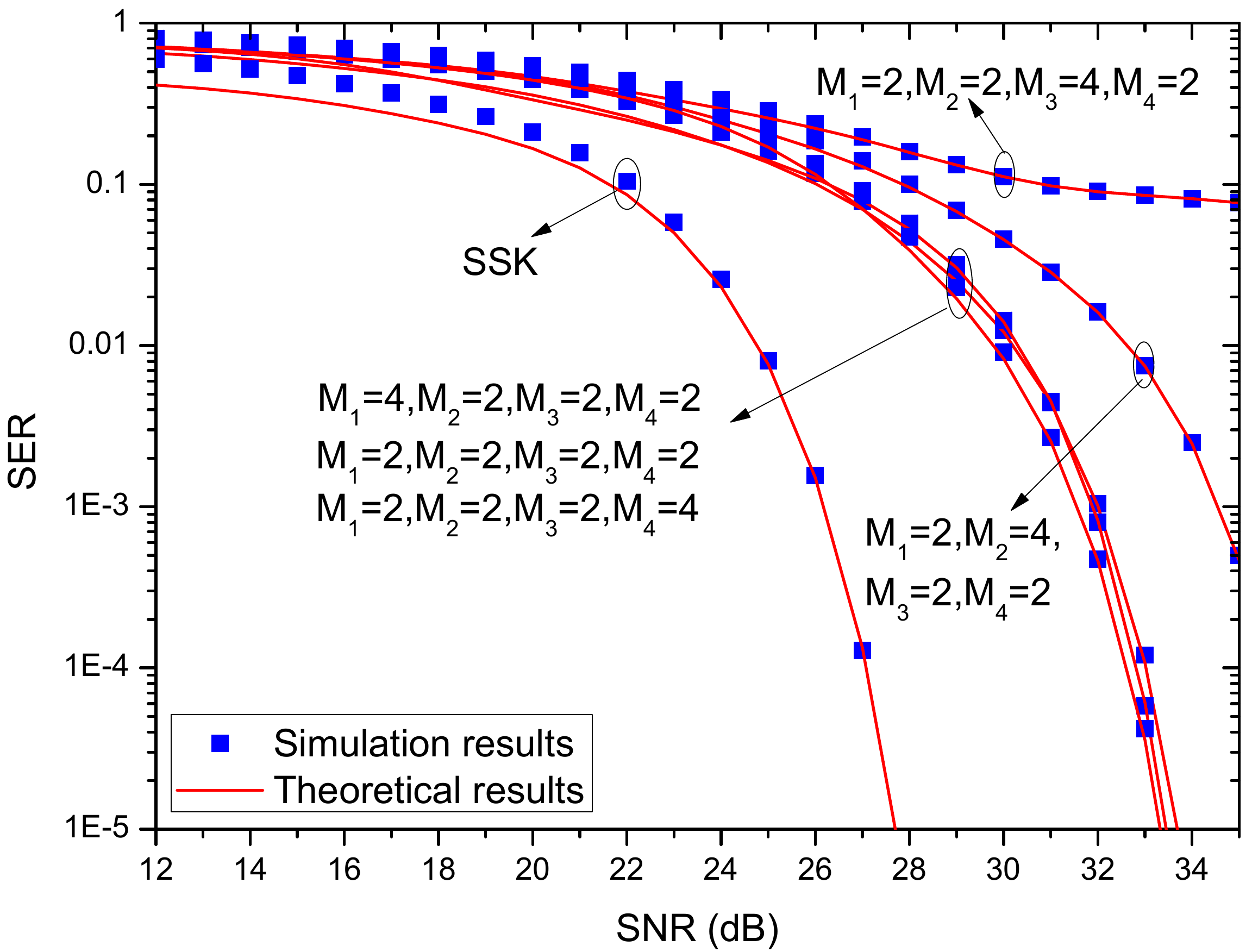}
\caption{SER versus SNR for the $4 \times 1$ VLC system under scenario 2}
\label{fig6}
\end{minipage}
\begin{minipage}[t]{0.5\linewidth}
\centering
\includegraphics[width=2.7in]{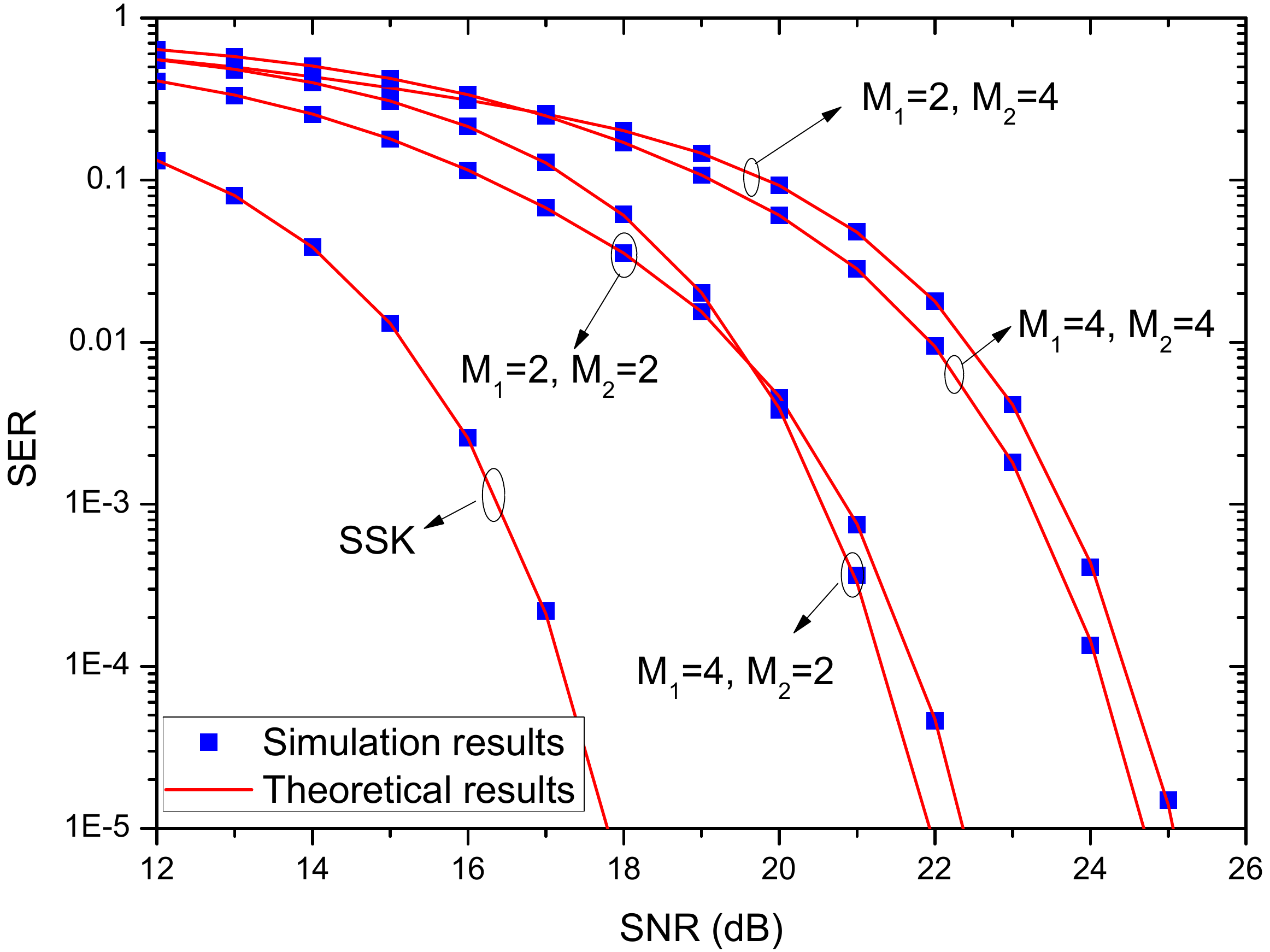}
\caption{SER versus SNR for the $2 \times 2$ VLC system under scenario 3}
\label{fig6_1}
\end{minipage}
\begin{minipage}[t]{0.5\linewidth}
\centering
\includegraphics[width=2.7in]{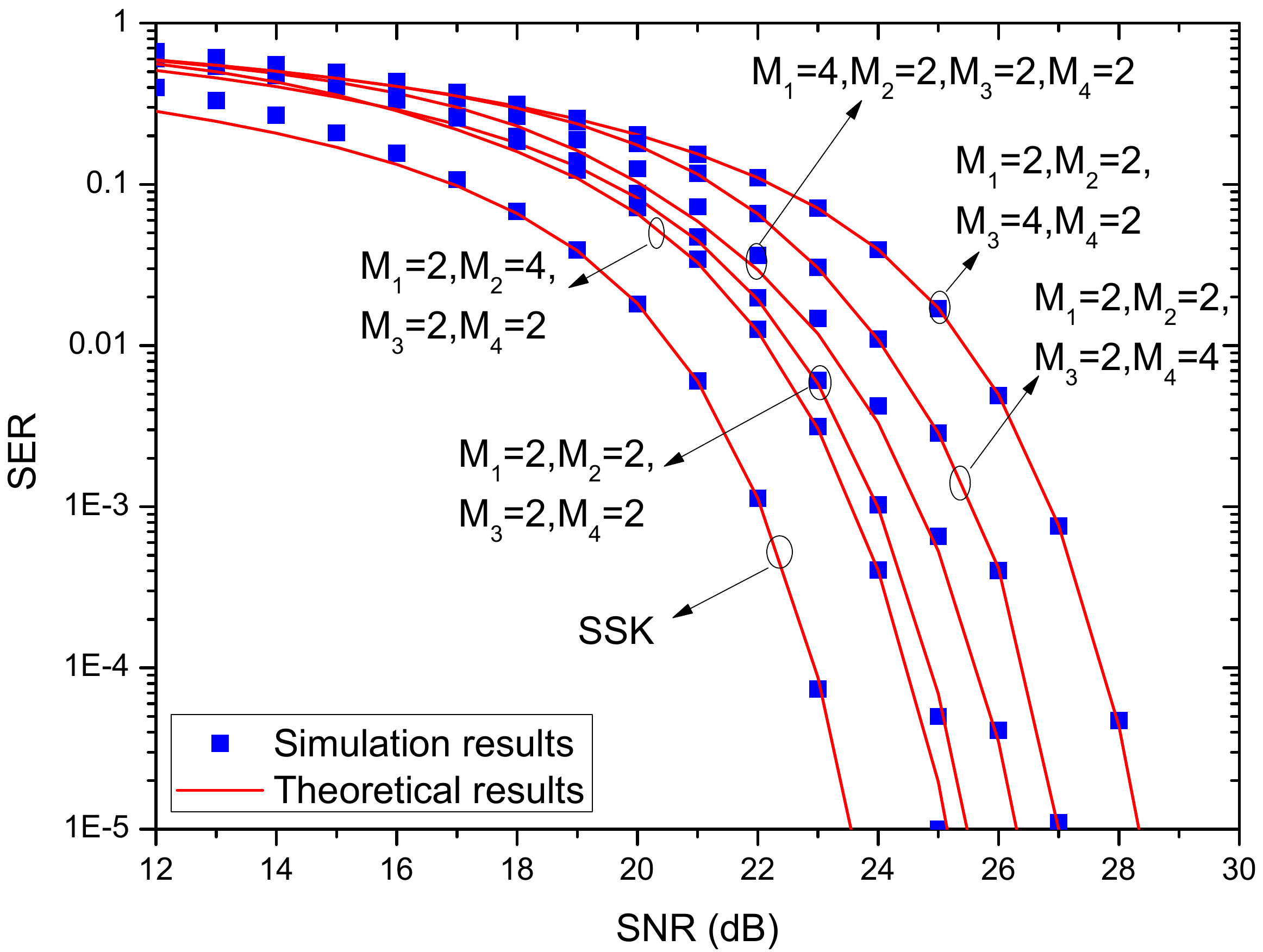}
\caption{SER versus SNR for the $4 \times 4$ VLC system under scenario 4}
\label{fig6_2}
\end{minipage}
\end{figure}

To verify the accuracy of the theoretical expression of the average SER,
Fig. \ref{fig5} - Fig. \ref{fig6_2} illustrate theoretical results and simulation results of the average SER under scenarios 1, 2, 3 and 4, respectively.
Obviously, the average SERs in Figs. \ref{fig5}-\ref{fig6_2} decrease with the increase of SNR.
In each figure, it can be seen that the derived theoretical values of the average SER match simulation results very well especially in the high SNR regime.
Moreover, the modulation orders on the LEDs have a large impact on the system performance.
That is, the systems with different modulation order combinations have different SER performance.
As can be seen in each figure, the system with the SSK scheme always achieves the best SER performance.
However, its spectral efficiency is the lowest one.
For scenario 1-scenario 4, the system with modulation order combination [2, 4], [2, 2, 4, 2],
[2, 4] and [2, 2, 4, 2] achieves the worst SER performance, respectively.
This indicates that it is very necessary to employ the ASM scheme to improve system performance.

%

\begin{figure}
\begin{minipage}[t]{0.5\linewidth}
\centering
\includegraphics[width=2.7in]{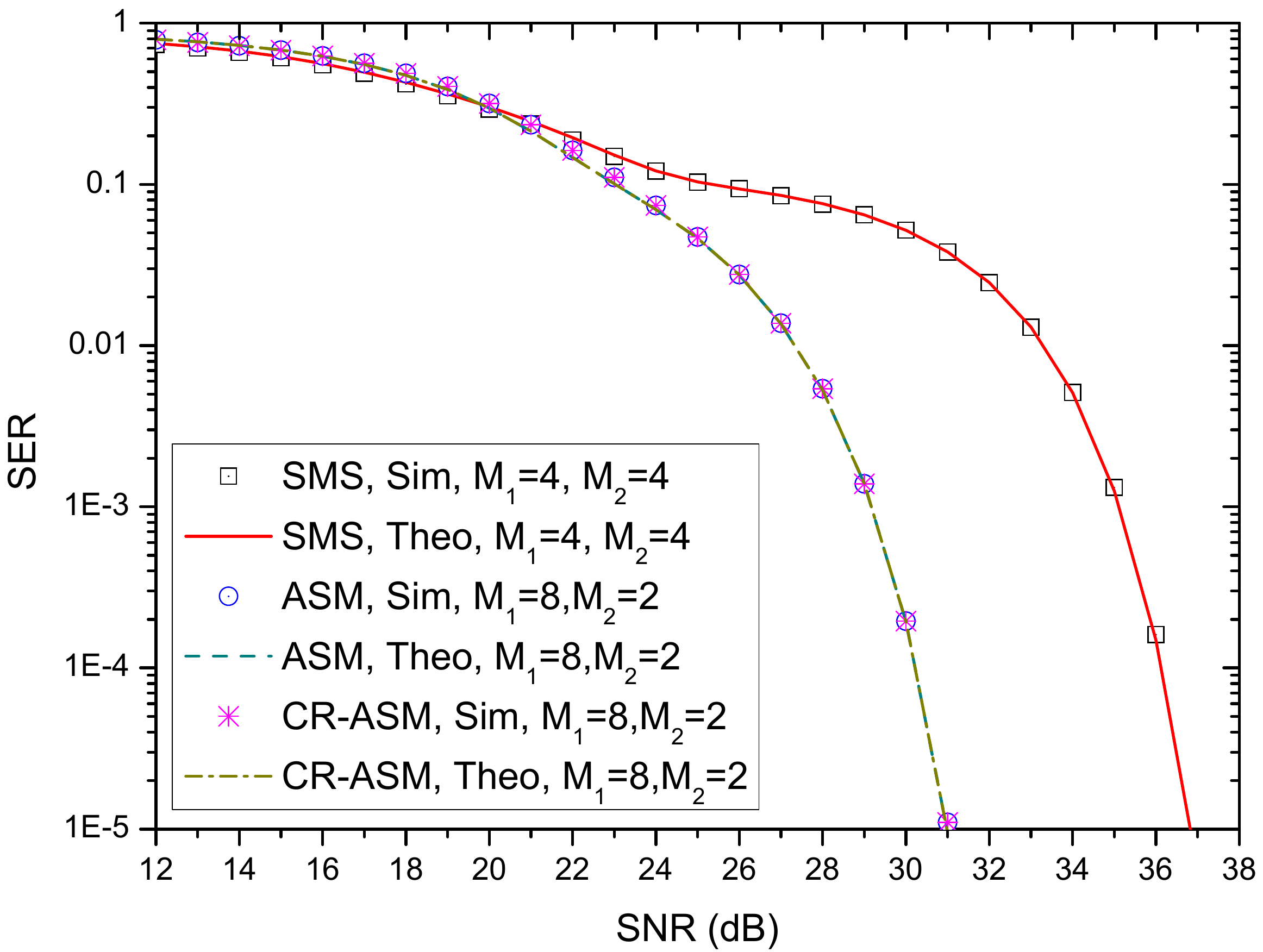}
\caption{SER comparisons for different schemes under scenario 1 with $m=3$ bit/s/Hz}
\label{fig7}
\end{minipage}
\begin{minipage}[t]{0.5\linewidth}
\centering
\includegraphics[width=2.7in]{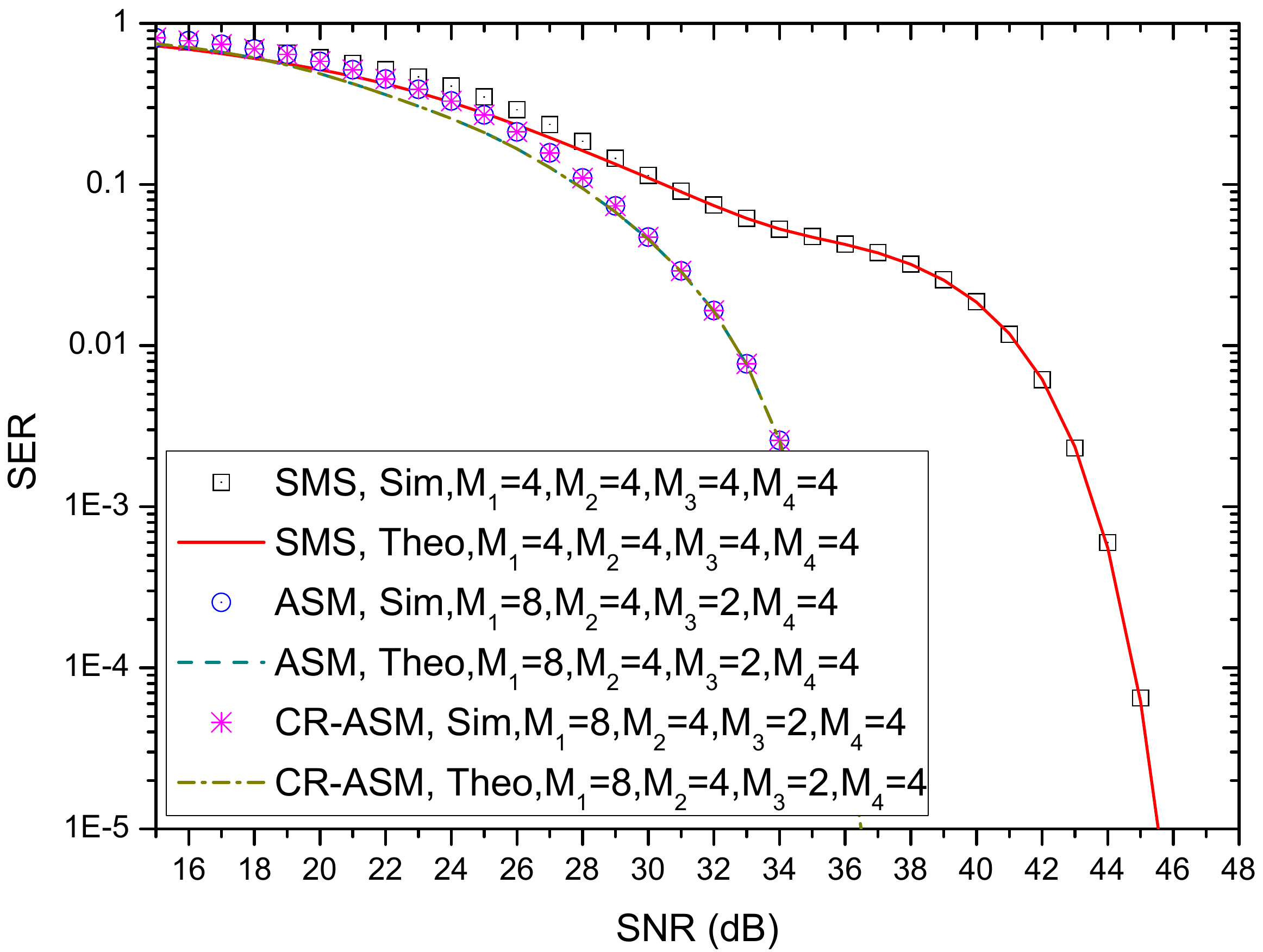}
\caption{SER comparisons for different schemes under scenario 2 with $m=4$ bit/s/Hz}
\label{fig8}
\end{minipage}
\begin{minipage}[t]{0.5\linewidth}
\centering
\includegraphics[width=2.7in]{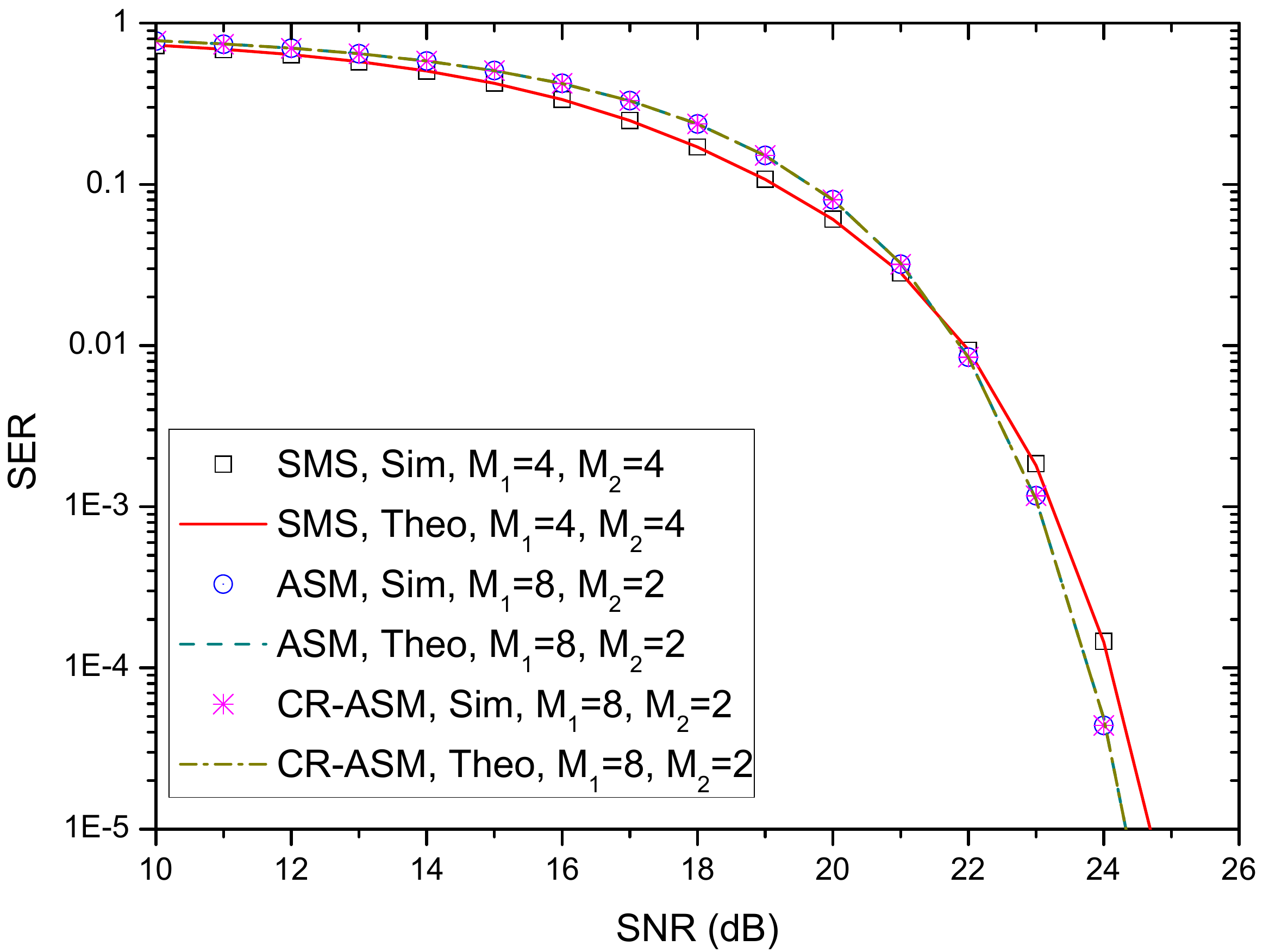}
\caption{SER comparisons for different schemes under scenario 3 with $m=3$ bit/s/Hz}
\label{fig8_1}
\end{minipage}
\begin{minipage}[t]{0.5\linewidth}
\centering
\includegraphics[width=2.7in]{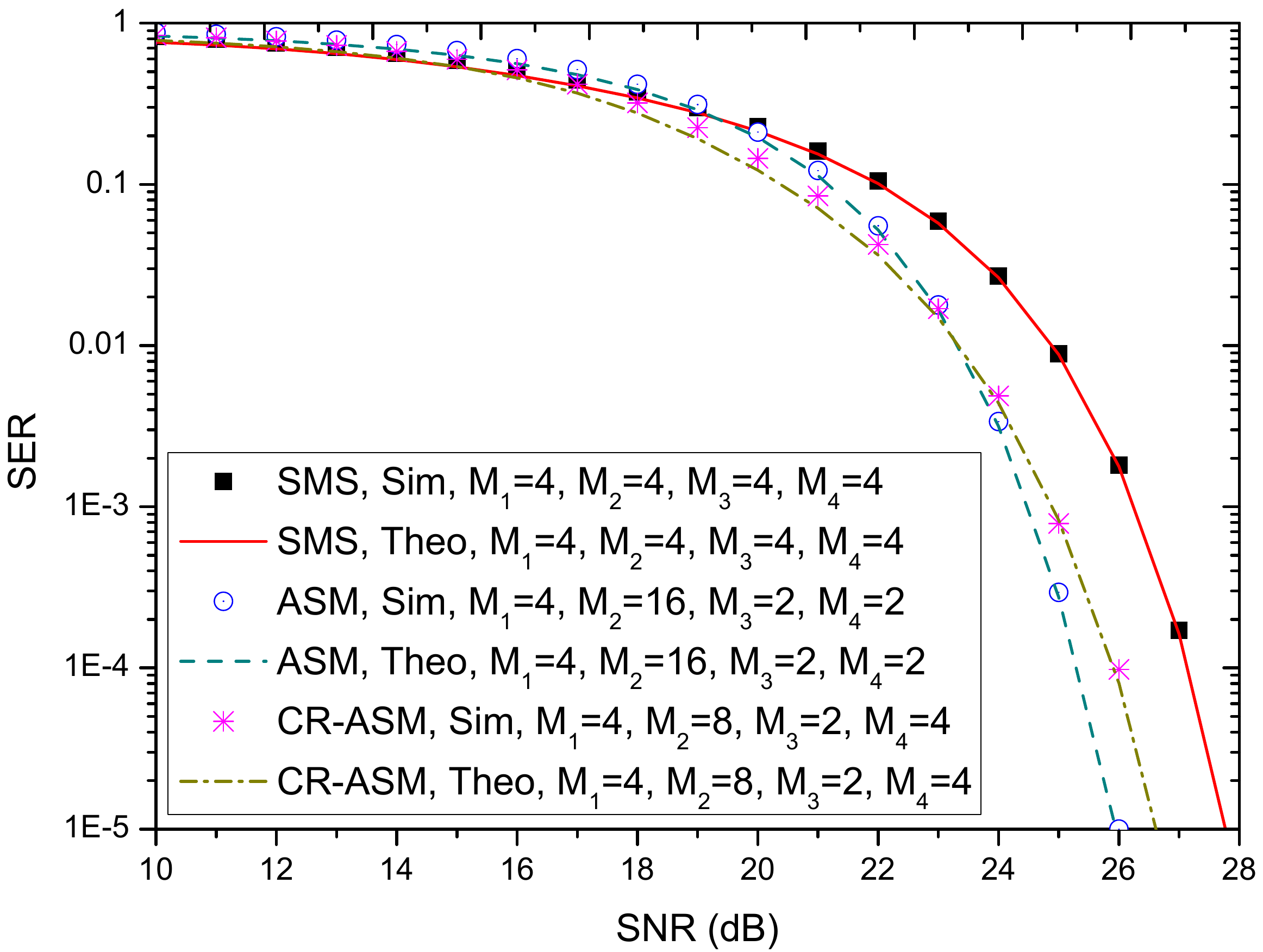}
\caption{SER comparisons for different schemes under scenario 4 with $m=4$ bit/s/Hz}
\label{fig8_2}
\end{minipage}
\end{figure}

Fig. \ref{fig7} - Fig. \ref{fig8_2} show the average SER performance comparisons for three space modulation schemes (i.e., ASM, CR-ASM, and SMS) under scenarios 1, 2, 3 and 4, respectively.
For all schemes, the theoretical values of the average SER match simulation results very well at high SNR, which verifies the accuracy of the derived expression of SER.
Note that the modulation order combinations of the ASM and the CR-ASM schemes are optimized by using Algorithms 1 and 2, respectively.
In Fig. \ref{fig7}, the spectral efficiency of the three space modulation schemes is the same and is equal to 3 bit/s/Hz.
The SMS scheme achieves the worst performance with modulation order on each LED is 4-PAM,
while the ASM scheme and the CR-ASM scheme achieve the same and better performance with the modulation order combination [8, 2] (i.e.,the first LED employs 8-PAM, the second LED employs 2-PAM).
In Fig. \ref{fig8}, the spectral efficiency of the three space modulation schemes is the same and is equal to 4 bit/s/Hz.
The SMS scheme achieves the worst performance with modulation order on each LED is 4-PAM,
and the ASM scheme and the CR-ASM scheme achieve the same and better performance with the modulation order combination [8, 4, 2, 4].
Although the ASM and the CR-ASM schemes achieve the comparable performance,
the CR-ASM scheme proposed in this paper is a better choice by considering the computational complexity.
Similar conclusions can also be found in Fig. \ref{fig8_1} and Fig. \ref{fig8_2}.
Comparing Fig. \ref{fig7} with Fig. \ref{fig8_1} (or comparing Fig. \ref{fig8} with Fig. \ref{fig8_2}), it can be observed that the SER performance improves with the increase of the number of PDs (i.e., $N_{\rm r}$) when the spectrum efficiency is fixed.

%

\section{Conclusion}
\label{section7}
In this paper, the theoretical expression of the average SER for the ASM-based VLC is derived to evaluate the system performance.
As special cases, the SERs for the VLC using SSK and SMS are also analyzed, respectively.
Numerical results indicate that the derived theoretical expression is quite accurate to evaluate system performance at high SNR.
To further improve the system performance, a modulation order optimization problem with fixed spectrum efficiency is formulated.
Two algorithms (i.e., the ASM scheme and the CR-ASM scheme) are proposed to solve the problem.
Numerical results indicate that the SSK scheme achieves the best performance. However, the signal constellation of the SSK scheme is not employed, and thus its spectral efficiency is low.
Moreover, the proposed two schemes obtain comparable performance,
and the proposed CR-ASM scheme can reduce the computational complexity.
Therefore, with a low computational complexity, the proposed CR-ASM scheme is a better choice for VLC.

\section*{Acknowledgements}
The authors wish to thank the anonymous reviewers for their valuable suggestions.


\bibliographystyle{IEEEtran}

\end{document}